\documentclass[10pt]{article}
\usepackage{graphicx}
\usepackage{amsmath,amssymb} 
\usepackage{amsthm,amsfonts}
\usepackage{color}

\newtheorem{theorem}{Theorem}

\allowdisplaybreaks
\newcommand{\R}{\mathbb{R}}
\newcommand{\e}{\mathrm{e}}
\DeclareMathOperator*{\argmin}{argmin}

\begin{document}

\title{3D Image Reconstruction from\\ X-Ray Measurements with Overlap} 

\author{Maria Klodt and Raphael Hauser \\ Mathematical Institute \\ University of Oxford}

\maketitle

\begin{abstract} 
3D image reconstruction from a set of X-ray projections is an important image reconstruction problem, with applications in medical imaging, industrial inspection and airport security. The innovation of X-ray emitter arrays allows for a novel type of X-ray scanners with multiple simultaneously emitting sources. However, two or more sources emitting at the same time can yield measurements from overlapping rays, imposing a new type of image reconstruction problem based on nonlinear constraints. Using traditional linear reconstruction methods, respective scanner geometries have to be implemented such that no rays overlap, which severely restricts the scanner design. We derive a new type of 3D image reconstruction model with nonlinear constraints, based on measurements with overlapping X-rays. Further, we show that the arising optimization problem is partially convex, and present an algorithm to solve it. Experiments show highly improved image reconstruction results from both simulated and real-world measurements.
\end{abstract}

\section{Introduction}

Reconstructing a three-dimensional image from a set of two-dimensional projections is a typical inverse problem.\let\thefootnote\relax\footnote{Certain aspects described in this publication have been filed for patent protection.}
\let\thefootnote\relax\footnote{Published in Computer Vision -- ECCV 2016. The final publication is available at 
link.springer.com/chapter/10.1007/978-3-319-46466-4\_2}
Applications of X-ray image reconstruction include medical imaging, industrial inspection and airport security. This paper addresses the problem of reconstructing a 3D image from a set of 2D X-ray projections with overlapping measurements. 
Traditionally, X-ray tomography is based on a single moving source following specific positions around the object or person to be scanned, for example the source moves around the reconstruction domain. In this type of set-up, images are taken sequentially, which implies that overlapping rays cannot occur. 
However, when using an emitter array, several sources can emit X-rays simultaneously \cite{Gonzales14,chen2015transmission}. This new type of X-ray scanning can lead to a new form of measurements: overlapping X-rays where rays from more than one emitter reach the same detector at the same time.  

Since measurements are usually undersampled, i.e.\ the number of measurements is less than the number of unknowns, prior information about the image to be reconstructed can help to improve the image reconstructions. Sparsity priors about the image are widely used. In compressed sensing research it has been shown that minimizing the L1 norm subject to linear constraints can yield sparse solutions for certain classes of reconstruction problems \cite{Candes06,Donoho06}. 
For medical images, the Total Variation (TV) norm has been shown to provide a suitable sparsity prior, e.g.\ for magnetic resonance imaging (MRI) \cite{Lustig07,Ma08cvpr} and X-ray computed tomography (CT) \cite{Yan2011}.
The arising reconstruction problem is closely related to another Computer Vision problem: 3D surface reconstruction from 2D photographs. Instead of X-rays, minimizing the TV norm on ray constraints can also be based on visual rays, e.g.\ to impose silhouette consistency in 3D multi-view reconstruction \cite{Kolev08}.
Other related applications include X-ray image reconstruction with shape priors \cite{Serradell11} and segmentation of CT reconstructions \cite{KimTB15}.

The relevant literature on sparse reconstruction mainly considers linear constraints. In most related work on nonlinear compressed sensing the nonlinearity is due to noise \cite{Ehleretal2014,cosamp}.  Other recent works considering nonlinearity in compressed sensing include a generalization to quasi-linear compressed sensing \cite{EhlerFS14} and compressed sensing with quadratic constraints \cite{LiVoroninski2012}. 
More recent works on compressed sensing consider more general constraints \cite{blumensath2012}, where a restricted isometry property for nonlinear measurements is defined as the distance to a linearized version.

\begin{figure}[t]
\begin{center}
\includegraphics{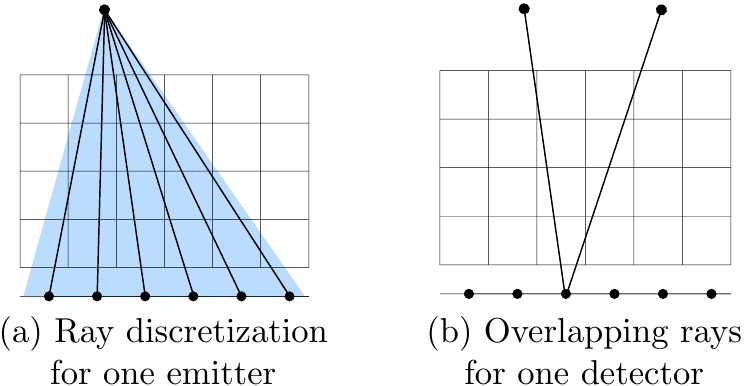}
\end{center}
\caption{Scanner set-up with a set of X-ray emitters above the reconstruction domain, and a flat panel detector below (schematic 2D view). (a): A finite set of rays is modeled for each emitter, connecting the emitter source positions with the center of each detector pixel, which discretizes the emitter cone (blue).
(b): If multiple sources emit simultaneously, it can occur that two or more rays reach the same detector at the same time, yielding a nonlinear overlapping measurement.}
\label{fig:rays}
\end{figure}

Measurements from overlapping rays yield a different type of nonlinearity. 
In a set-up where multiple sources (partially) emit simultaneously, using image reconstruction from linear constraints requires the scanner to be designed such that the emitter cones do not overlap. This restriction to small emitter cones and small stand-off distances (distances between emitters and reconstruction domain) highly constrains the scanner geometry. In particular, hand-held devices where the distance from sources to detectors cannot be positioned exactly are constrained to unnecessarily small collimation angles, in order to guarantee that no rays overlap. An appropriate handling of overlap is therefore essential to improve robustness with respect to varying emitter and detector positions. It allows for larger emitter collimation angles of the device, which is necessary for accurate 3D image reconstructions.

In this paper an image reconstruction problem arising from measurements with overlapping X-rays is derived, and an algorithm to solve it is presented.
The main contributions are:
\begin{itemize}
\item derivation of a new model for image reconstruction from X-ray measurements with overlap, which arises from a novel type of multi-source X-ray scanners,
\item proof that the arising optimization problem is partially convex,
\item a novel optimization method to solve 3D X-ray image reconstruction from measurements with overlap,
\item demonstrated practicability of the method by comparison to ground truth from simulated data, as well as practicability under real-world conditions using real X-ray measurements. Experiments show both quantitatively and visually improved results compared to traditional linear reconstruction.
\end{itemize}
 
The outline of the paper is as follows: Sec.\ \ref{sec:linearconstraints} briefly reviews X-ray image reconstruction from sequential measurements, yielding linear constraints. Sec.\ \ref{sec:nonlinear} presents a novel formulation of an image reconstruction problem based on nonlinear constraints from overlapping rays. The section further presents an algorithm to solve the arising optimization problem. Sec.\ \ref{sec:implementation} gives details on the implementation. Sec.\ \ref{sec:experiments} shows experiments with overlapping X-rays based on simulated as well as real-world measurements. Sec.\ \ref{sec:conclusion} concludes the paper.


\section{X-Ray Image Reconstruction with Linear Constraints}
\label{sec:linearconstraints}

The scanner set-up considered in this paper consists of a set of emitters which are located above the region to reconstruct and a rectangular panel of detector pixels located below the region. The X-rays emitted from the conoidal sources are discretized as a finite set of rays connecting each detector pixel center with the emitter source positions. 
We further assume a three-dimensional region of interest in which the object to reconstruct is located, discretized to a Cartesian grid of $n$ voxels. A schematic two-dimensional view of the rays for one emitter is depicted in Fig.\ \ref{fig:rays} (a).

Assuming homogenuous material per voxel, no ray scattering, no noise, and no overlap, the {\it{Beer-Lambert}} law decribes the relation of material properties and radiation intensity $I_{Ej}$ at the emitter and radition intensity $I_{Dj}$ at the detector:
\begin{equation}
I_{Dj} = I_{Ej} \exp \left( \sum_{i=1}^n -\xi_{ij} x_i \right) 
\label{eq:beerlambert}
\end{equation}
for measurements $j=1,\dots,m$. Here, $\xi_{ij}$ is the distance that the ray corresponding to measurement $j$ traverses through voxel $i$.
Reformulation of \eqref{eq:beerlambert} yields a sparse linear system of equations
\begin{equation}
\log\left(\frac{I_{Dj}}{I_{Ej}}\right) =  \sum_{i=1}^n -\xi_{ij} x_i 
~\Leftrightarrow~ Ax=b
\end{equation}
where $A\in\R^{m\times n}$ with $a_{ij}=-\xi_{ij}$ is a matrix projecting the 3D image $x$ to the 2D projections $b$, depending on scanner geometry and grid discretization. The measurements are stacked to vector $b\in\R^m$ with $b_j=\log\left(I_{Dj}/I_{Ej}\right)$, and the (unknown) densities $x_i$ per voxel are stacked to vector $x\in\R^n$. Hence, each column of $A$ corresponds to a voxel index $i$, and each row corresponds to a measurement $j$. Furthermore, $A$ is sparse, because in general, each ray intersects only a small number of voxels, implying that most of the entries $a_{ij}$ are zero.

Usually, the measurements are undersampled, i.e. $m<\!\!<n$, and prior information about the unknown $x$ can help to recover the solution. In the context of compressed sensing one is interested in the sparsest solution which can be represented by the reconstruction problem
\begin{equation}
\min \|x\|_1 \quad\text{s.t.}\quad Ax=b.
\label{eq:linear-constraints}
\end{equation}
It has been shown that for certain matrices, minimizing the L1 norm can yield sparse solutions for $Ax=b$, i.e.\ $x$ with few non-zero elements \cite{Candes06,Donoho06}.


\section{Nonlinear Measurements with Overlapping Rays}
\label{sec:nonlinear}

The standard X-ray image reconstruction model \eqref{eq:linear-constraints} based on linear constraints is obtained from sequential exposures of X-ray sources, i.e.\ no rays overlap. However, if multiple X-ray sources emit simultaneously, it can occur that two or more rays reach the same detector at the same time (see Fig.\ \ref{fig:rays} (b)). We will see that the measurements of overlapping rays are not linear anymore, however we can show that they are convex. In this section we will derive a convex formulation of the arising reconstruction problem, and propose a forward-backward splitting algorithm to optimize it.


\subsection{A Model for X-Ray Image Reconstruction with Overlap}

If two rays reach the same detector at the same time the intensity at measurement $j$ sums up to
\begin{equation}
\frac{I_{Dj}}{I_{Ej}} =
\exp \left( \sum_{i=1}^n -\xi_{ij1} x_i \right) + 
\exp \left( \sum_{i=1}^n -\xi_{ij2} x_i \right)
\end{equation}
where each of the two terms on the right-hand side corresponds to the attenuation along one ray, and equal emitter intensities $I_{Ej}$ are assumed for the simultaneously active emitters.
A more general formulation for $p$ rays simultaneously reaching measurement $j$ is given by
\begin{equation}
\frac{I_{Dj}}{I_{Ej}} = \sum_{k=1}^{p} \exp \left( \sum_{i=1}^n -\xi_{ijk} {x}_i \right) :=\psi_j(x)
\label{eq:measurements-overlap}
\end{equation}
where the coefficients $\xi_{ijk}$ correspond to the length of the intersection between voxel $i$ and the ray from emitter $k$ to the detector corresponding to measurement $j$. 
Assuming again that we are interested in a sparse reconstruction $x$, minimization with L1 prior yields 
\begin{equation}
\min_x \|x\|_1 {\quad{\rm{s.t.}}\quad} \sum_{k=1}^{p} \exp \left( \sum_{i=1}^n -\xi_{ijk} {x}_i \right) = b_j , \quad\forall j=1,\dots,m
\label{eq:overlap-l1-constraint}
\end{equation}
with measurements $b_j={I_{Dj}}/{I_{Ej}}$.
The coefficients $-\xi_{ijk}$ can be represented with sparse vectors $r_{jk}\in\R^n$:
\begin{equation}
r_{jk}=\begin{pmatrix} -\xi_{1jk}, \dots, -\xi_{njk} \end{pmatrix}^\top ,~ j=1,\dots,m, k=1,\dots,p.
\end{equation}
Allowing for noise in the data constraint term, we propose the following least squares formulation of \eqref{eq:overlap-l1-constraint}:
\begin{equation}
\min_x \left\lbrace\|x\|_1 + \frac{1}{2\mu} \sum_{j=1}^m\left( \sum_{k=1}^{p} \exp\left(\sum_{i=1}^n -\xi_{ijk} x_i\right) - b_j \right)^2 \right\rbrace 
\label{eq:overlap-l1-ls}
\end{equation}
with regularization parameter $\mu>0$ which provides a balance between sparsity prior and data fidelity term.
The formulation \eqref{eq:overlap-l1-ls} corresponds to an optimization problem of the form
\begin{equation}
\min_x \left\lbrace f(x) + g(x) \right\rbrace
\label{eq:optimization-general}
\end{equation}
with convex non-differentiable $f:\R^n\rightarrow\R$:
\begin{equation}
f(x) = \|x\|_1
\end{equation}
and partially convex, differentiable $g:\R^n\rightarrow\R$:
\begin{equation}
g(x) = \frac{1}{2\mu} \sum_{j=1}^m\left( \sum_{k=1}^{p} \exp\left(\sum_{i=1}^n -\xi_{ijk} x_i\right) - b_j \right)^2.
\end{equation}

\begin{theorem}
Let $\hat{x}$ be a minimizer that fulfils the measurements \eqref{eq:measurements-overlap}. The function $g(x)$ is partially convex for $0\le x\le\hat{x}$.
\label{th:convex}
\end{theorem}
\begin{theorem} 
The gradient $\nabla g$ is Lipschitz continuous for all $x\ge0$ with Lip\-schitz constant $L=\frac{1}{\mu}2mp^2 \xi_{max}^2$. Here, $\xi_{max}$ denotes the maximum value of all $\xi_{ijk}$:
\begin{equation} 
\xi_{max} := \max\left\lbrace \xi_{ijk} \,|\, i\in\lbrace 1,\dots,n\rbrace, j\in\lbrace1,\dots,m\rbrace,k\in\lbrace1,\dots,p\rbrace\right\rbrace
\end{equation}
which is bounded by the maximum intersection length of a ray with a voxel. For a voxel size $h$ this is $\xi_{max}\le\sqrt{3}h$. 
\label{th:lipschitz}
\end{theorem}
\noindent Proofs for Theorems \ref{th:convex} and \ref{th:lipschitz} are provided in the appendix.


\subsection{Optimization with Forward-Backward Splitting}

An optimization problem of the form \eqref{eq:optimization-general} can be solved using the first-order forward-backward splitting update sequence for $t=0,1,2,\dots$ and $x^0=0$:
\begin{equation}
x^{t+1} = {\mathrm{prox}}_{\lambda f}(x^t-\lambda \nabla g(x^t))
\label{eq:update-sequence}
\end{equation}
with convergence rate ${\mathcal{O}}(1/t)$ and step size $\lambda=1/L$ \cite{Combettes2005}. 
Note that due to the generally high dimension of the reconstruction problem, second-order optimization is infeasible, as the Hessian matrix has dimension $n^2$.
It has been shown in \cite{Combettes2005} that the update sequence \eqref{eq:update-sequence} converges to a minimum of \eqref{eq:optimization-general}, if $f$ is a lower semicontinuous convex function, and $g$ is convex, differentiable and has a Lipschitz continuous gradient.
The proximal operator ${\mathrm{prox}}_{\lambda f}:\R^n\rightarrow\R^n$ is defined as
\begin{equation}
{\mathrm{prox}}_{\lambda f}(x) = \argmin_u \left\lbrace f(u) + \tfrac{1}{2\lambda}\|u-x\|^2_2\right\rbrace.
\end{equation}
For the L1 ball $f(x)=\|x\|_1$ it can be computed analytically and is given by the ``soft thresholding operator'' \cite{donoho1998}:
\begin{equation}
{\mathrm{prox}}_{\lambda f}(x) = 
\begin{cases}
x-\lambda, &  x>\lambda \\
0, & |x| \le \lambda \\
x+\lambda, &  x<-\lambda \\
\end{cases}
\end{equation}
For a recent overview of proximal operators and algorithms, see for example \cite{Parikh2014}.

For an optimization of \eqref{eq:overlap-l1-ls}, the initialization of $x$ should be smaller than a minimizer $\hat{x}$, because $g$ is partially convex for $x\le\hat{x}$. Knowing that the densities $x$ cannot be negative, we chose the lower bound $x^0=0$. Furthermore, the step sizes $\lambda$ have to be chosen such that $x\le\hat{x}$ is assured for all $t$.
We determine the step sizes $\lambda$ using a backtracking line search algorithm \cite{BeckTeboulle2009}, while here we have to ensure the constraint $x\le\hat{x}$. Although in general, $\hat{x}$ is unknown, the measurements $b$ are given with $b_j=\psi_j(\hat{x})$, and hence we constrain the line search by 
\begin{equation}
\psi_j(x) \ge b_j {\text{ for all }} j=1,\dots,m.
\end{equation}
We propose the following forward-backward splitting algorithm to minimize \eqref{eq:overlap-l1-ls}:
\begin{enumerate}
\setlength{\itemsep}{3pt}
\item[]\hskip-10pt {\bf{Input:}}\\ $b\in\R^m$ : vector of measurements\\ $r_{jk}\in\R^n$ : sparse vectors of intersection lengths of rays and voxels \\
$c\in(0,1)$ : line search control parameter\\
$L=\tfrac{1}{\mu}2mp^2\xi_{max}^2$ : Lipschitz constant for $g$
\item[]\hskip-10pt {\bf{Initialize}} $x^0=0$.
\item[]\hskip-10pt {\bf{Iterate}} for $t=0,1,2,\dots$:
\item[] {\it{1. Compute search direction:}} 
\begin{equation}
\nabla g(x^t) = \frac{1}{\mu}\sum_{j=1}^m \left(\sum_{k=1}^p \e^{r_{jk}^\top x^t} -b_j \right)
\left( \sum_{k=1}^p r_{jk} \e^{r_{jk}^\top x^t} \right)
\end{equation}
\item[] {\it{2. Backtracking line search:}}
\begin{eqnarray}
&&\lambda = 1/L\\
&&x^{new} = {\mathrm{prox}}_{\lambda f}(x^t-\lambda \nabla g(x^t))\\
&&\text{while } \exists j\in\lbrace1,\dots,m\rbrace:\psi_j(x^{new}) < b_j :\nonumber \\
&&\qquad\lambda\leftarrow\lambda c \\
&&\qquad x^{new} = {\mathrm{prox}}_{\lambda f}(x^t-\lambda \nabla g(x^t))
\end{eqnarray}
\item[] {\it{3. Update $x$:}}  ~~$x^{t+1}=x^{new}$.
\end{enumerate}


\section{Implementation}
\label{sec:implementation}

To compute the projection vectors $r_{jk}$, the intersection lengths $\xi_{ijk}$ of each voxel with each ray have to be computed. The choice of an efficient algorithm is crucial to the performance of the method to avoid a computation time of $O(mn)$ for $m$ rays and $n$ voxels. Efficient methods that can be used include line rastering algorithms used for ray tracing in computer graphics applications. We implemented the algorithm of \cite{Amanatides87} which is based on traversing only those voxels whith non-zero intersection length for each ray.

As regularization term we implemented the discrete version of the isotropic Total Variation (TV) norm, which has been shown to provide a suitable sparsity prior for X-ray image reconstruction \cite{Lustig07}. On a discrete three-dimensional grid it is given by:
\begin{equation}
\|x\|_{TV} = \sum_{i,j,k} |(\nabla x)_{i,j,k}|
\end{equation}
with
\begin{equation}
|(\nabla x)_{i,j,k}| = \frac{1}{h}
\sqrt{(x_{i,j,k}-x_{i+1,j,k})^2 + (x_{i,j,k}-x_{i,j+1,k})^2 + (x_{i,j,k}-x_{i,j,k+1})^2} \nonumber
\end{equation}
where $h$ denotes the spacing in the grid, and here $i,j,k$ denote the indices of the discrete locations in the image domain.


\section{Experiments with Overlapping X-Rays}
\label{sec:experiments}

This experimental section aims to investigate how overlapping of X-rays affects the reconstruction accuracy, and which amount of overlap is tolerable in order to be able to recover the 3D image. The proposed method is tested for both simulated data and real X-ray measurements, which allows to both draw conclusions about reconstruction errors where a ground truth is available, and observing the behaviour of the method under real-world conditions including noise.

Since this paper is based on a novel type of reconstruction problem, we compare to a reconstruction based on non-overlapping measurements only, where we eliminate all measurements $b_j$ and corresponding vectors $r_{jk}$ where the number of rays per measurement is $\ge2$. This non-overlapping case yields a reconstruction problem of type \eqref{eq:linear-constraints}, and can be optimized using a standard image reconstruction algorithm based on linear constraints \cite{ChambollePock2011}.

The distance $d(x)$ of a reconstruction $x$ to the reference $\hat{x}$ is measured by 
\begin{equation} d(x) = {\|x-\hat{x}\|_2}/{\|\hat{x}\|_2}. \label{eq:distance} \end{equation}


\begin{figure}[t]
\begin{center}
\includegraphics{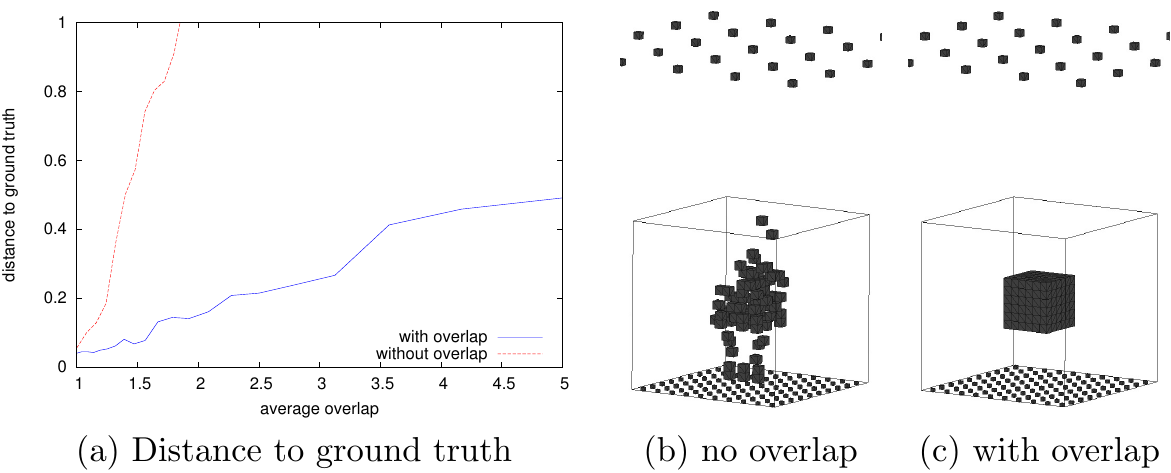}
\end{center}
\caption{Reconstruction of a cube from simulated data. (a): Reconstruction error measured as distance to ground truth for increasing amount of overlap $\bar{p}$, i.e.\ the average number of rays per measurement. (b),(c): Reconstruction and scanner geometry for average overlap of $\bar{p}\approx2$. Emitter positions are shown above the reconstruction domain, detectors below. (b): reconstruction from non-overlapping measurements only, (c): reconstruction with overlap.}
\label{fig:cube-rec}
\end{figure}

\subsection{Reconstruction from Simulated Data}

Fig.\ \ref{fig:cube-rec} (a) shows reconstruction errors for simulated measurements of a three-dimensional cube, comparing the proposed method with overlap to the linear reconstruction from non-overlapping measurements only. 
The x-axis shows increasing amount of average overlap $\bar{p}$ which we define as the average number of rays per measurement, i.e.\ the average number of rays reaching the same detector pixel at the same time. 
We compute measurements with overlap by varying the number of exposures, i.e.\ the number of simultaneously emitting sources. Overlapping rays then result from randomly chosen simultaneously active emitters. Hence, an overlap of $\bar{p}=1$ corresponds to sequential exposures. The rest of the experimental set-up is not changed during the experiment.
The y-axis shows the relative distance $d(x)$ of the reconstruction $x$ to the ground truth $\hat{x}$, computed by \eqref{eq:distance}.
For increasing amount of overlap, the reconstruction from non-overlapping constraints is based on fewer measurements, resulting in an empty volume reconstructed. Thus, for $x\approx0$ the plot reaches $d(x)\approx1$, which can be seen in the graph.
While both methods yield increasing reconstruction errors for increasing amount of overlap, the proposed method with handling of overlap yields significantly smaller reconstruction errors compared to reconstructions from non-overlapping measurements only. 

The scanner geometry used for this experiment is visualized in Fig.\ \ref{fig:cube-rec} (b) and (c): The reconstruction domain of 20$\times$20$\times$20 voxels is visualized by the grid bounding box, above, 5$\times$5 emitters are located with a stand-off distance of 20 voxels, and directly below the domain, a panel of 10$\times$10 detectors.
The binary test object $\hat{x}$ consists of cube of density $\hat{x}_i=1$ and size 6$\times$6 located in the center of the domain. Outside the cube, the test object has density $\hat{x}_i=0$.
Inside the bounding boxes are thresholded versions of the reconstructed densities, i.e.\ all voxels with $x_i\ge0.5$, that have been reconstructed from the set-up with average overlap of $\bar{p}\approx2$. Fig.\ \ref{fig:cube-rec} (b) shows the reconstruction from non-overlapping measurements only, using optimization with linear constraints. Fig.\ \ref{fig:cube-rec} (c) shows the reconstruction from all measurements, including overlap, using the proposed method.
The results show that also by visual comparison, the method with overlap yields a clearly better reconstruction.


\subsection{Reconstruction from Real-World Measurements}

The experiments with real X-ray measurements have been taken using a similar general scanner set-up as the simulated data: Above the reconstruction domain, a rectangular array of X-ray emitters is located, and directly below, a flat panel detector of 512$\times$512 pixels and size 13$\times$13 cm$^2$.
Two different test objects were measured and reconstructed: The data set ``letters'' consists of two wooden letters, stacked on top of each other, and measurements were taken from 14$\times$13=182 equally spaced emitter positions. The data set ``polecat'' is the skull of a polecat, measured from 13$\times$9=117 different emitter positions.
Overlapping rays are simulated from the sequential measurements by adding detector images of randomly chosen emitter positions. Hence, the intensities of two or more different sources are accumulated in the measurements.

\begin{figure}[t]
\begin{center}
\includegraphics{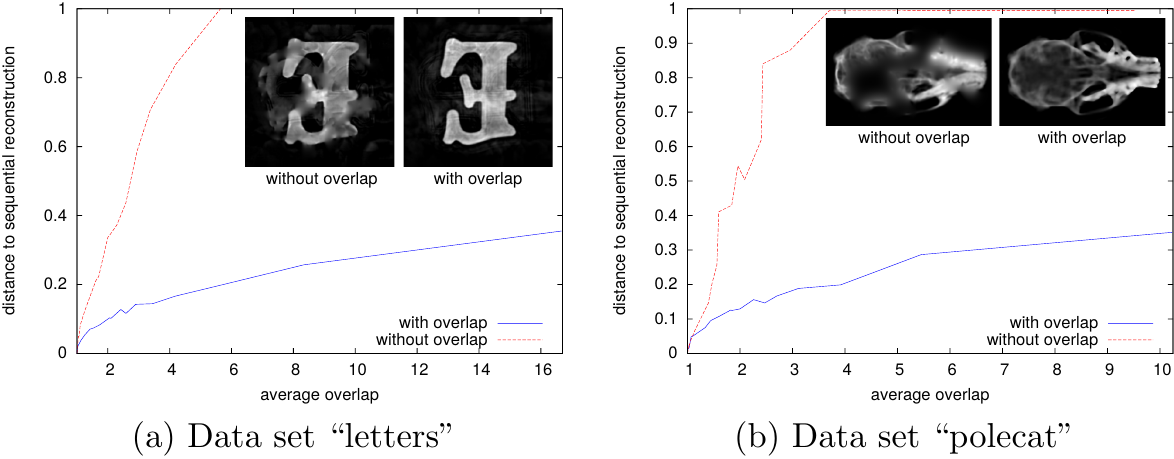}
\end{center}
\caption{Reconstruction from real X-ray measurements, for the two data sets ``letters'' and ``polecat''. Reconstruction errors for increasing amount of overlap are measured as distance to the respective reconstruction from sequential measurements. The images in the top right corners show reconstructions for an average overlap of $\bar{p}\approx2$, comparing linear reconstruction from non-overlapping measurements to the proposed method with handling of overlap.}
\label{fig:comparison}
\end{figure}

Fig.\ \ref{fig:comparison} (a) and (b) show the distance to the reconstruction from sequential measurements for increasing amount of overlap. The reconstruction with overlap is compared to the reconstruction which rejects the overlapping measurements, for the two data sets. 
All reconstructions are computed from the same geometry of emitters, detectors and grid dimension. The only parameter that is varying during one experiment is the number of simultaneously emitting X-ray sources, which determines the average amount of overlap per measurement.
Again, the x-axis shows the average amount of overlap per measurement, while the y-axis shows the distance to the reconstruction from sequential measurements, measured by \eqref{eq:distance}. Since a ground truth is not available for the real data, we consider the respective reconstruction from sequential exposures as reference $\hat{x}$, which corresponds to a reconstruction with overlap of $\bar{p}=1$. 

The two graphs allow to observe how much the reconstruction quality decreases with increasing amount of overlap. As in the case of simulated measurements, again we observe that increasing amount of overlap increases reconstruction errors, while the proposed method to handle overlapping rays clearly outperforms the reconstruction from non-overlapping measurements only. 
This is also visible in the images of reconstructed densities shown in the figure. The images show one slice of the reconstructed 3D densities, for an average overlap of $\bar{p}\approx2$, using measurements from non-overlapping measurements only, and the proposed method with overlap.


\begin{figure}[t]
\begin{center}
\includegraphics{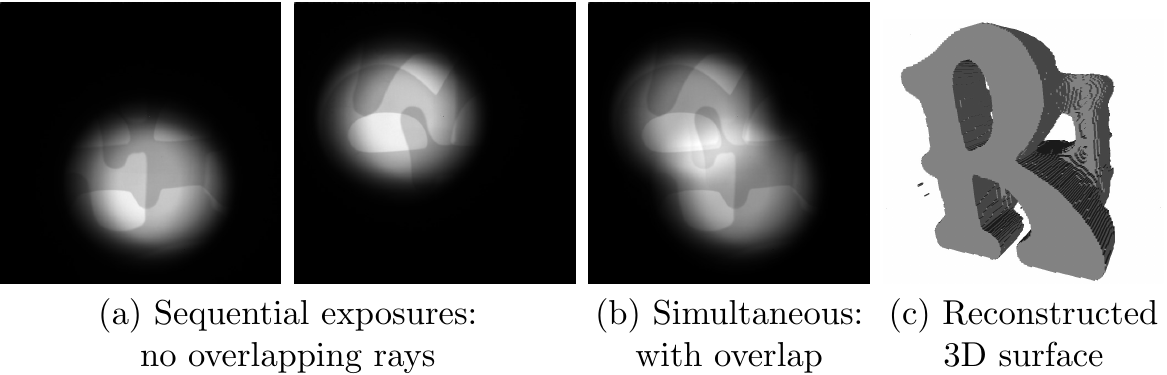}
\end{center}
\caption{Two of 182 detector images from sequential (a) and simultaneous (b) exposures. Scanned were two stacked wooden letters, images taken from above. (c): 3D view of reconstructed densities, thresholded at $x\ge0.035$.}
\label{fig:letters-input}
\begin{center}
\includegraphics{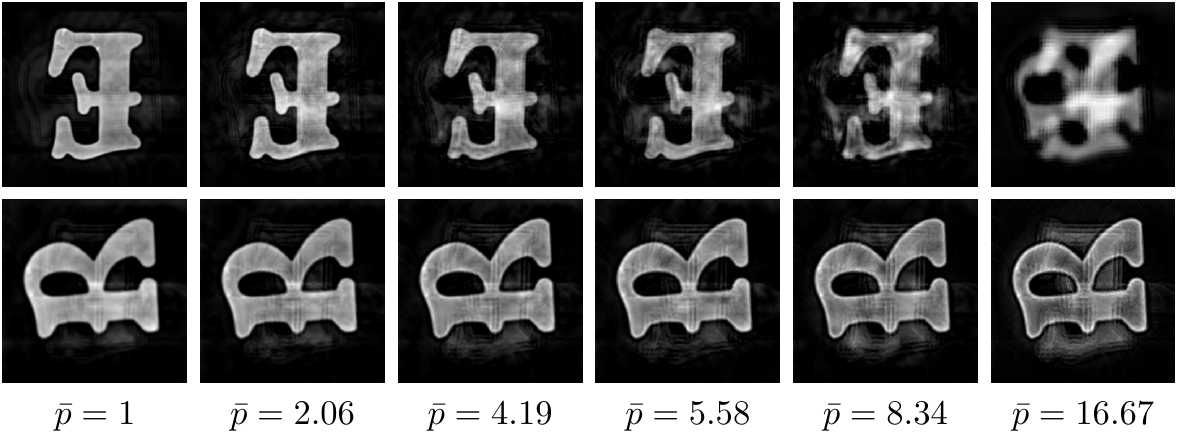}
\end{center}
\caption{Slices of reconstructed densities for the data set ``letters'', using varying amount of average overlap $\bar{p}$. First row: slice 5 of 20, second row: slice 19 of 20.}
\label{fig:letters-overlap}
\end{figure}

Fig.\ \ref{fig:letters-input} (a) shows two of the 182 detector images measured for the data set ``letters''. The two images can be added to obtain overlapping measurements, shown in Fig\ \ref{fig:letters-input} (b). 
Due to the limited field of view of each emitter cone, not all measurements in Fig\ \ref{fig:letters-input} (b) are overlapping. Only a part of the image, where the two emitter cones intersect, contains measurements with two overlapping rays, the other pixels have overlap of 1. Note that pixels outside any of the emitter cones are considered not visible and are excluded from the measurements.
Fig.\ \ref{fig:letters-input} (c) shows the 3D surface of reconstructed densities, thresholded at $x\ge0.035$, computed from  sequential measurements, i.e.\ overlap of 1.

Fig.\ \ref{fig:letters-overlap} shows two different slices of the reconstructed densities, for increasing amount of average overlap $\bar{p}$. The first image with overlap of $\bar{p}=1$ corresponds to sequential exposures.
The reconstructions were computed on a voxel volume of dimension 512$\times$512$\times$20. While the x- and y-axes have the same dimension as the detector panel, we chose a significantly smaller resolution of the z-axis, because it corresponds to the main orientation of rays, which implies reduced recoverability along the z-axis. The slices shown in the figure are along the z-axis, and thus parallel to the detector panel.
Note that the reconstruction is capable to clearly seperate the two letters from each other in the different slices, although they are overlayed in all of the projections (see Fig.\ \ref{fig:letters-input}).
The experiment shows that up to an average overlap of $\bar{p}\approx2$, the method is capable to recover almost the same image quality compared to sequential exposures.


\begin{figure}[t]
\begin{center}
\includegraphics{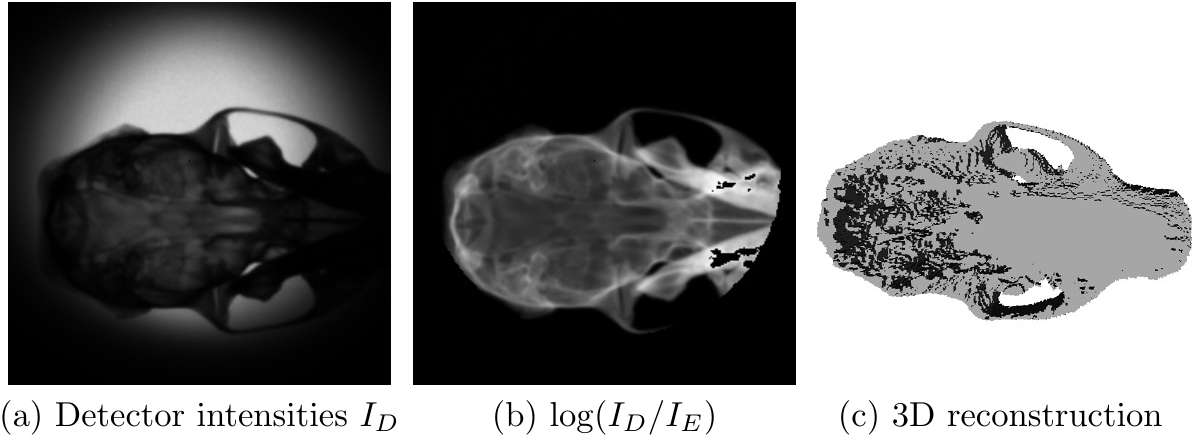}
\end{center}
\caption{Data set ``polecat''. (a): Measurements at the detector for one of 117 views. (b): Log-ratio of detector and emitter intensities. (c): 3D surface of reconstructed densities, thresholded at $x\ge0.2$.}
\label{fig:polecat-input}
\begin{center}
\includegraphics{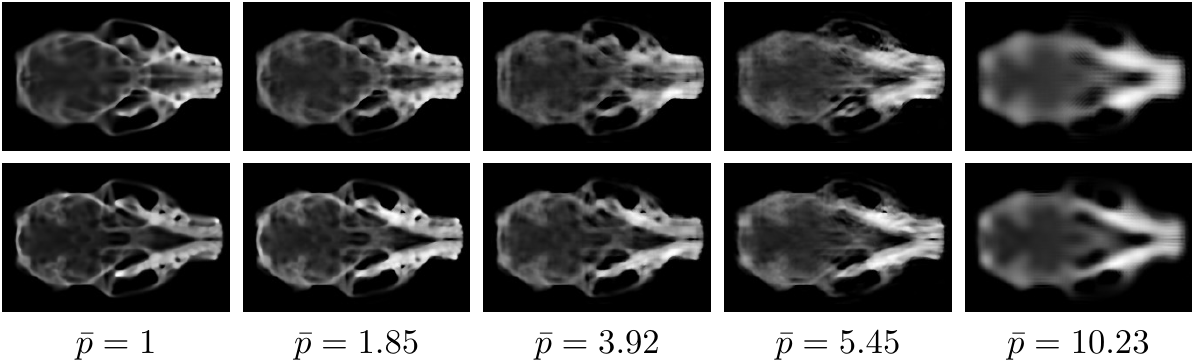}
\end{center}
\caption{Reconstructed densities of a polecat skull with increasing amount of average overlap $\bar{p}$. First row: slice 9 of 20, second row: slice 12 of 20. }
\label{fig:polecat-overlap}
\end{figure}

Fig.\ \ref{fig:polecat-input} (a) shows measurements at the detector for one of the 117 images of the ``polecat'' data set. Fig.\ \ref{fig:polecat-input} (b) shows the logarithm of the ratio of emitter and detector intensities. The dark spots on the right part of the visible area correspond to points where the detector measures zero intensity. Possible reasons for no measurements include defective detector pixels, and low emitter intensity in combination with dense material which attenuates all the radiation.
Fig.\ \ref{fig:polecat-input} (c) shows the 3D surface of reconstructed densities, thresholded at $x\ge0.2$, computed from  sequential measurements, i.e.\ overlap of 1.

Fig.\ \ref{fig:polecat-overlap} shows two different slices of the 3D reconstructed densities of the ``polecat'' data set. The grid consists of 512$\times$512$\times$20 voxels, and the overlap of $\bar{p}=1$ corresponds to sequential exposures. 
Again, the slices are along the horizontal axis, i.e. parallel to the detector panel and vertical to the main direction of rays.
The results show that, similar to the observation of the ``letters'' data set, an average overlap of $\bar{p}\approx2$ is tolerable to recover almost the same image quality compared to sequential exposures. 


\subsection{Computation Time}

Reconstruction from linear constraints is faster than the proposed reconstruction with handling of overlapping rays. One of the reasons is that step sizes for the linear constraints can be computed analytically, while the convex constraints need a backtracking line search to compute step sizes. Furthermore, the computation of the exponential function in the data constraints of overlapping rays is time consuming. In our experiments, the method with overlap is approximately 5 times slower than the corresponding reconstruction from non-overlapping measurements, which we consider a tolerable factor given the highly improved reconstruction results.

In medical applications, the computation time for the reconstruction is generally not a limiting factor, but the total exposure time is, as the patient is moving. This is where the method with overlap has major advantages, as in certain cases the time budget does not allow for taking sufficiently many sequential exposures to acquire data without overlap.


\section{Conclusions}
\label{sec:conclusion}

This paper has presented a practicable method to solve a novel type of optimization problem arising from overlapping X-rays. The method is based on modelling nonlinear constraints in sparse reconstruction. A new type of image reconstruction problem has been formulated, based on nonlinear constraints from overlapping X-ray measurements. 
Moreover, it has been proved that the arising optimization problem is partially convex. 
Results have shown that the proposed method improves reconstruction accuracies compared to linear constraints, using both simulated and real-world measurements. Experiments show that the method is capable to recover nearly the same image quality compared to sequential exposures up to an average overlap of $\sim$2.

The proposed method has a potential impact on X-ray scanner designs, because it allows to handle measurements from overlapping X-rays, where traditional X-ray image reconstruction from linear constraints cannot cope with. This novelty highly increases robustness with respect to stand-off distances and emitter collimation angles, allowing now for a new class of hand-held X-ray scanning devices. Furthermore it allows for more flexibility in the design of X-ray scanner geometry.

\section*{Acknowledgements} We thank Adaptix Ltd for providing the X-ray measurements used in the experiments. This work was supported by Adaptix Ltd and EPSRC EP/K503769/1.


\begin{appendix}
\section{Proof of Theorems}

\paragraph{Proof of Theorem \ref{th:convex} (convexity of $g$)}
Let $\psi(x)$ defined as in \eqref{eq:measurements-overlap}.
The convexity of 
\begin{equation}
\psi_j(x) = \sum_{k=1}^{p} \exp \left( \sum_{i=1}^n -\xi_{ijk} {x}_i \right)
\end{equation}
follows from convexity of the exponential function.
For all $x,y\in\R^n$ with $0\le x_i\le\hat{x_i}$ and $0\le y_i\le\hat{x_i}$ for $i=1,\dots,n$ and $\lambda\in[0,1]$
\begin{eqnarray}
\hskip-15pt g(\lambda x + (1-\lambda)y)\hskip-5pt &=& 
\frac{1}{2\mu} \sum_{j=1}^m \left(\psi_j(\lambda x+(1-\lambda)y)-b_j\right)^2
\label{eq:proof_convexity1}\\
&\le& \frac{1}{2\mu} \sum_{j=1}^m \left(\lambda\psi_j(x)+(1-\lambda)\psi_j(y)-b_j\right)^2
\label{eq:proof_convexity2}\\
&\le& \lambda\frac{1}{2\mu} \sum_{j=1}^m \left(\psi_j(x)\!-\!b_j\right)^2 +
(1\!-\!\lambda)\frac{1}{2\mu} \sum_{j=1}^m \left(\psi_j(y)\!-\!b_j\right)^2
\label{eq:proof_convexity3}\\
&=& \lambda g(x) + (1-\lambda)g(y)
\end{eqnarray}
where \eqref{eq:proof_convexity2} follows from the monotonicity of the square function for positive values, and \eqref{eq:proof_convexity3} follows from convexity of $\psi_j$ for all $j=1,\dots,m$.
\hfill$\Box$

\paragraph{Proof of Theorem \ref{th:lipschitz} (Lipschitz continuity of $\nabla g$)}
We will show $|g^{\prime\prime}|\le L$ for all $x_i\ge0$ with $i=1,\dots,n$ which implies Lipschitz continuity of $\nabla g$.
The first and second derivatives of $g$ are given by:
\begin{equation}
\frac{\partial g}{\partial x_{i_1}} = \frac{1}{\mu}\sum_{j=1}^m \left(\sum_{k=1}^p \exp\left(\sum_{i=1}^n -\xi_{ijk} x_i\right) -b_j \right) \!
\left( \sum_{k=1}^p -\xi_{{i_1}jk} \exp\left(\sum_{i=1}^n -\xi_{ijk} x_i\right)\!\right) 
\end{equation}
\begin{eqnarray}
&&\hskip-10pt\frac{\partial^2 g}{\partial x_{i_1}\partial x_{i_2}}= 
\frac{1}{\mu}\sum_{j=1}^m \left[
\sum_{k=1}^p -\xi_{{i_2}jk} \exp\left(\sum_{i=1}^n -\xi_{ijk} x_i\!\right) \!
 \sum_{k=1}^p -\xi_{{i_1}jk} \exp\left(\sum_{i=1}^n -\xi_{ijk} x_i\!\right) \right. \nonumber \\
&&\hskip-10pt +
\left.
\left(\sum_{k=1}^p \exp\left(\sum_{i=1}^n -\xi_{ijk} x_i\right) -b_j \right) \!
\left( \sum_{k=1}^p \xi_{{i_1}jk} \xi_{{i_2}jk} \exp\left(\sum_{i=1}^n -\xi_{ijk} x_i\right)\!\right)\!\right]. 
\label{eq:second-derivative}
\end{eqnarray}
The upper bound $L=\frac{1}{\mu}2mp^2 \xi_{max}^2$ on $g^{\prime\prime}$ derives from $\xi_{ijk}\in[0,\xi_{max}]$ as follows:
\begin{eqnarray}
\hskip-15pt\left|\frac{\partial^2 g}{\partial x_{i_1}\partial x_{i_2}}\right|
&=&
\left|\frac{1}{\mu}\sum_{j=1}^m \left(
\left( \sum_{k=1}^p -\xi_{{i_2}jk} \e^{r_{jk}^\top x}\right) \!
\left( \sum_{k=1}^p -\xi_{{i_1}jk} \e^{r_{jk}^\top x}\right) \right.\right. \nonumber \\
&&+
\left.\left.
\left( \sum_{k=1}^p \e^{r_{jk}^\top x} -\sum_{k=1}^p \e^{r_{jk}^\top \hat{x}} \right) 
\left( \sum_{k=1}^p \xi_{{i_1}jk} \xi_{{i_2}jk} \e^{r_{jk}^\top x}\right)\!\right)\right|
\label{eq:proof_lipschitz1}
\\
&\le&
\frac{1}{\mu}\sum_{j=1}^m \left(
\left| \sum_{k=1}^p -\xi_{{i_2}jk} \e^{r_{jk}^\top x}\right| \!
\left| \sum_{k=1}^p -\xi_{{i_1}jk} \e^{r_{jk}^\top x}\right| \right. \nonumber \\
&&+
\left.
\left| \sum_{k=1}^p \e^{r_{jk}^\top x} -\sum_{k=1}^p \e^{r_{jk}^\top \hat{x}} \right| 
\left| \sum_{k=1}^p \xi_{{i_1}jk} \xi_{{i_2}jk} \e^{r_{jk}^\top x}\right|\right)
\label{eq:proof_lipschitz2}
\\
&\le&
\frac{1}{\mu}\sum_{j=1}^m \left(
\left( \sum_{k=1}^p \xi_{{i_2}jk}\right) \!
\left( \sum_{k=1}^p \xi_{{i_1}jk}\right) +
p \left( \sum_{k=1}^p \xi_{{i_1}jk} \xi_{{i_2}jk} \right)\!\right)
\label{eq:proof_lipschitz3}
\\
&\le&
\frac{1}{\mu}\sum_{j=1}^m  2p^2 \xi_{max}^2 = \frac{1}{\mu}  2mp^2 \xi_{max}^2 :=L.
\label{eq:proof_lipschitz4}
\end{eqnarray}
while \eqref{eq:proof_lipschitz2} follows from the triangle inequality and \eqref{eq:proof_lipschitz3} follows from the fact that $\xi_{ijk}\ge0$ and $\exp(r_{jk}^\top x)\in(0,1]$.
\hfill$\Box$


\end{appendix}

\bibliographystyle{plain}
\bibliography{references}
\end{document}